# Automated System for Improving RSS Feeds Data Quality

JOAN FIGUEROLA HURTADO, Edinburgh Napier University

Nowadays, the majority of RSS feeds provide incomplete information about their news items. The lack of information leads to engagement loss in users. We present a new automated system for improving the RSS feeds' data quality. RSS feeds provide a list of the latest news items ordered by date. Therefore, it makes it easy for a web crawler to precisely locate the item and extract its raw content. Then it identifies where the main content is located and extracts: main text corpus, relevant keywords, bigrams, best image and predicts the category of the item. The output of the system is an enhanced RSS feed. The proposed system showed an average item data quality improvement from 39.98% to 95.62%.

Categories and Subject Descriptors: **H.3.3 [Information Search and Retrieval]**: Selection Process

## 1. INTRODUCTION

The Internet has become a platform for mass collaboration in content production and consumption. There are an increasing number of people consuming their daily news via online sources. Online newspapers have developed significant presence in the Internet.

A data feed is a mechanism for users to receive updated data from data sources. News feed is a popular form of data feed. They are known as RSS (Rich Site Summary) and are used by news sites to publish information about their articles. It is widely used by publishers as it enables them to syndicate data automatically. A standard XML file format ensures compatibility with many different machines. Data feeds usually require structured data. At the present time unstructured data, e.g. HTML pages, dominate the Web. As a result, data feeds have the potential to make a bigger impact on the Web in the future. Items compose a feed as shown in Figure 1. Feed and items have specific attributes that describe each entity.

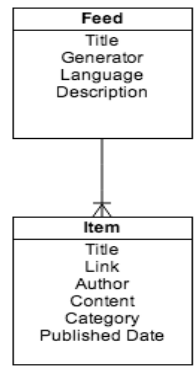

Fig. 1. RSS feed structure.

However, RSS items do not always provide all attributes or complete information. In some specific cases, publishers deliberately choose to just provide an excerpt of the article followed by a "continue reading" link in order to send the user to the publisher site. These are the most common issues:

- Content only has three or less sentences instead of full text
- Articles have no categories
- Images have low resolution
- Main image is missing
- Author is missing

## 2. RELATED WORK AND MOTIVATION
### 2.1 Related work

Some approaches use supervised learning to extract the news article content wrapper from the general news pages. Shinnou et al. (2004) gave an extraction wrapper learning method and expected to learn the extraction rules which could be applied to news pages from other various news sites. Reis et al. (2004) gave a calculation of the edit distance between two given trees for the automatic Web news article contents extraction. Fukumoto et al. (2002) focused on subject shift and presented a method for extracting key paragraphs from documents that discuss the same event. It uses the results of event tracking which starts from a few sample documents and finds all subsequent documents. However, if a news site uses too many different layouts in the news pages, the learning procedure can be highly time consuming and negatively affect the precision of the algorithm [F. Fukumoto et al. 2002].

Webstemmer is a Web crawler and HTML layout analyser that automatically extract main text from a news site without having advertisements and navigation links mixed up [Y. Shinyama 2009]. However, this approach runs slowly when parsing and extracting contents, and sometimes news titles are missing [Y. Shinyama 2009]. TSReC [Y. Li et al. 2006] provides a hybrid method for news article contents extraction. It uses tag sequence and tree matching to detect the parts of a news article [H. Shinnou et al. 2001].

Han et al. (2009) gave an automation web news article content extraction system based on RSS feeds. Firstly, it extracts the news title and URL of each news page from RSS feeds. Secondly, the news title is split to get a keyword list and use them to detect the position of news title in the news page. Then, paragraphs of the news article are recognised by news title position and keyword list. Finally, all the paragraphs of news article contents are extracted out of the full text of the news page. The main weaknesses of this approach are:

- It requires having the news title and its location within the page
- News titles may be so short that two or more similar sentences are found in different locations within the page. It makes it difficult to recognise where the main content is
- There is no news title
- There is just one possible news title, but is not the real news title (Related Content)

Embedly is a web extraction tool and API that extracts an accurate list of attributes from a single web page [Embedly Core Development Team 2014]. However, it does not provide a system to effectively crawl news sites as we explain in the section 3.2. In addition, it does not classify articles or extract the most relevant bigrams. Lastly, it uses a proprietary technology, which makes it difficult to evaluate.

**2.2 Motivation**

As stated before, items with attributes compose news feeds. Each item should have all the attributes so news aggregators can get the most out of it. Hence, a quality item should have a creative and descriptive title alongside the main content. There is no restriction on the length of the content so the full text is always recommended. Following that, an item should also have a category, author and published date. It helps classifying items and makes them searchable. Lastly, items should have compelling images optimized for web and mobile devices [GoDataFeed Editorial Team 2013].

The main issue with previous work is that it focuses on extracting the main content of the article, thus there might be important attributes missing such as: category, thumbnail, author, published date and keywords. These attributes are sometimes equally or more important than the main content as it allows news aggregation sites to organize news and make them searchable in order to offer a compelling user experience. Besides that, the accuracy of previous mentioned systems depends upon the page layout, thus unusual changes might lead to poor accuracy.

The main aim of this research is to improve the quality of data feeds; particularly RSS feeds because they are widely used by online publishers. A study shows that 97% of America's top online newspapers have RSS feeds [The Bivings Group 2007]. They provide a rich variety of content that can be easily collected. However, findings also apply into other data feeds such as e-commerce product feeds or data APIs.

Two major objectives are to be achieved: the first intends to define a set of heuristics to compute a numeric value for the quality of each item. The second objective aims to design and develop a system to improve the quality of news feeds.

The proposed system differentiates from related work as follows:
- It does not only provide an automated method for extracting the main content of news pages, it also provides automated methods to extract or predict main attributes such as: category, content text, thumbnails and keywords.
- It uses an asynchronous crawler suitable for high performance environments.
- It has an RSS feed as input and an enhanced RSS feed as output making it really easy to use for RSS applications.

**3. METHODOLOGY**

The methodology chapter intends to explain the methods used to develop the proposed system. Firstly, a set of heuristics is defined to assess the quality of items. Then, the methods to collect RSS feeds are detailed. Finally, the set of methods used to improve the data of each item attribute are explained.

**3.1 Heuristics To Asses Item Quality**

Items' quality needs to be measured in order to measure the performance of the proposed system. To compute the quality of an item, we assign a weight to each attribute/feature. The item quality is computed as follows:

$$IQ(x) = (T\ x\ 15 + C\ x\ 51 + A\ x\ 2 + D\ x\ 2 + C\ x\ 15 + I\ x\ 15) / 100 \quad (2)$$

Each feature takes a binary value (1 or 0) depending on whether they exist in the item or not. $T$ is the title of the original article, $A$ is the original author, $D$ is the original published date, $C$ is the original category and $I$ is the original Image. $C$ is the content of the article and it takes value 1 if the content length of the item extracted by the proposed system is greater than the original content length.

A quality item should have a creative and descriptive title. There is no restriction on the length of the main content, thus the full text is always recommended. Following that, an item should also have a category, author and published date. It makes items searchable and easy to classify. Furthermore, items should have compelling images optimized for web devices [GoDataFeed Editorial Team 2013].

The weights for each feature in the formula are chosen according to the definition of quality item and feedback given by publishers.

It is necessary to point out that there might be different definitions of data quality depending on the needs of the user of the system. For instance, an image aggregation application might consider images more relevant than content. Therefore, $I$ variable in the IQ formula could be changed to 51 and the C variable could be changed to 15.

### 3.2 Collection

Two methods to collect information from RSS feeds and HTML pages are covered in this section.

#### 3.2.1 Feed Parser

*Feedparser* is an open source library that parses RSS feeds in Python [Feedparser Core Development Team 2011]. Feeds and item's attributes can be accessed once the RSS is parsed. RSS feeds must comply with a specification [Harvard Editorial Team 2001]. Generally, this method is used to collect the initial items from RSS feeds before the quality of them is improved.

#### 3.2.2 Web Crawler Guided By A RSS Feed

A crawler guided by a RSS feed has been developed to extract information from articles. The RSS feed provides a list of the latest news items ordered by date. Therefore, It makes it easy for the crawler to know where to go to extract the latest information about articles. Adam, G et al. proposed a system named *AdvaRSS* and state that, "The difference between *AdvaRSS* and a usual crawler is the fact that the news is produced in a random order any time of the day and thus the crawling mechanism has to be efficient in order to obtain each news article that occurs in a news site immediately after its publishing." [Patras Bouras et al. 2009]

The link attribute from the news feed is used to collect the raw HTML content of items using the developed crawling system. It has been developed on top of *Scrapy* [*Scrapy* Core Development Team 2009], which is an asynchronous crawling framework written with *Twisted* [Twisted Core Development Team 2009], Twisted is a popular event-driven networking framework for Python.

The main reason to choose *Scrapy* as opposed of other crawling systems is that its asynchronous nature makes it suitable for high performance environments. Furthermore, it is an open source project widely and successfully used in a variety of large-scale applications.

### 3.3 Content Extraction

The location of the main content needs to be identified amongst all HTML tags to extract it. Firstly, the number of characters per tag is counted as shown in the first plot in the Figure 2. A pattern occurs in news pages; the main content is located in the region that has the highest number of characters per tag and is consecutive. Tags outside the main content region are filtered out as shown in the second plot in Figure 2, thus the main content is obtained. Finally, striping HTML tags from the main content retrieves the text corpus.

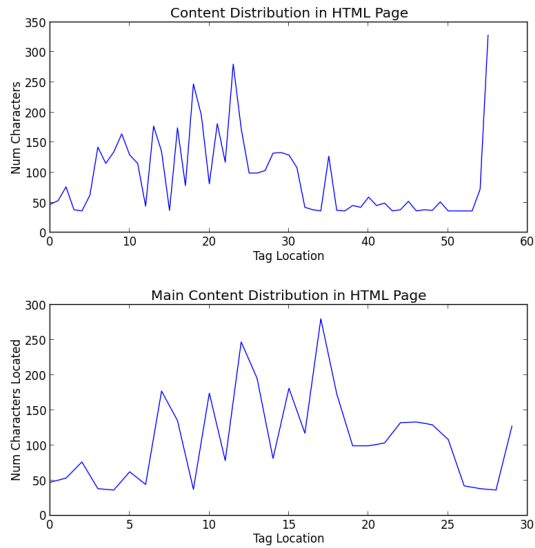

Fig. 2. Content Distribution.

### 3.4 Keyword Extraction

The Natural Language Toolkit (NLTK) is used to process the text corpus extracted in section 3.3 and 'mine' the most relevant keywords and bigrams. Firstly, the text corpus is tokenized by words. Irrelevant words commonly known as 'stopwords` are removed from the text corpus. After that, a frequency distribution function is applied to the tokenized corpus. As a result, a dictionary with *"<'keyword' : 'count'>"* is obtained.

A bigram is a pair of consecutive written units such as letters, syllables, or words [Oxford Dictionary Online 2014]. A bigram finder and counter method is applied to the tokenized corpus to find the most relevant bigrams. As a result, a dictionary *"<'bigram' : 'count'>"* is obtained. Top keywords and bigrams dictionaries are merged together and ranked by the count value resulting into a new dictionary.

### 3.5 Image Extraction

The main image will be searched within the crawled page. All the images are collected and the area of each one is calculated. After that, their area ranks them and the top image is selected as the main image of the item.

In some cases, there are no images in the page meaning that a less compelling user experience will be offered to end users of applications. Therefore, suitable images optimised for web devices are searched in public image databases using keywords previously extracted in section 3.4.

Another potential issue is that images might have low resolution; therefore manipulations on them such as scaling up will distort the image. Bad resolution images are replaced for high resolution ones optimized for web devices using the system mentioned in the previous paragraph.

**3.6 Item Classification**

It is necessary to clarify the difference between categories and keywords. Categories are broad and general, high level views that allow the user to easily identify a particular group of documents (or images). [Murat 2011] Keywords are more specific, low-level descriptions, offering a narrower view of whatever the subject of the document (or image) might be. [Murat 2011] For example:

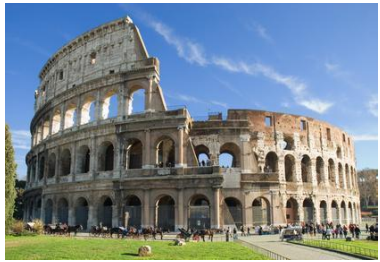

Fig. 3. Roman Coliseum. [A. Segre 2012]

Murat (2011) categorise Fig. 3 as follows:

- Architecture > Historical Architecture > Ruins
- Travel > European Landmarks > Rome

Murat (2011) describes Figure 3 using the following keywords: ancient, arches, architectural, architecture, arena, coliseum, gladiators, historic, Italian, monument Rome, roman, stone, tourism and travel.

**3.6.1   Supervised Multiclass Classification Model**

A machine learning multiclass classification model is trained using a machine-learning library called *Scikit-Learn*. It is built on top of the popular Python's packages: *Scipy*, *Numpy* and *Matplotlib*. The algorithm classifies articles into suitable classes. A one-vs-rest (OVR) or one-vs-all (OVA) [C. Bishop 2006] algorithm is used for reducing the problem of supervised multiclass classification to multiple binary classification problems. OVR requires: a learner or estimator, training data and expected results.

80K articles were crawled from public news outlets. They were labeled according the section they were in the news outlet. For example, articles under the technology section are classified as *science and technology*. The IPTC taxonomy for news is used to label the dataset and train the classification algorithm. There are 1150 categories and 5 levels of depth in the IPTC taxonomy for news [IPTC 2011]. However, only the 21 parent categories are used because the accuracy decreases dramatically as more children categories are added to the classification algorithm. The reason behind such behavior is the lack of labeled data samples.

Different estimators are benchmarked. A linear support vector machine estimator provided the highest accuracy and performance, thus it is the most suitable one for the model. As a result, classes are predicted with an average 80% accuracy.

**3.7 Author**

The author of the article is extracted from the HTML Meta tag author located at the header of the HTML page. If there is no author Meta tag, online news sites have profile pages for each member of the editorial team. Individual articles usually reference author's profile page using HTML links. Several online news sites tend to reference authors using this URL patterns:

*http://www.example.com/author[s]/[author_name]*
*http://www.example.com/people/[author_name]*
*http://www.example.com/user[s]/[author_name]*
*http://www.example.com/editor[s]/[author_name]*

Therefore, all individual article links were filtered according author's URL pattern in order to identify author's profile page URL. Then, the URL is parsed and the last part of the URL is beautified. To beautify and URL means to transform a part of an URL into readable text. For example, *christopher-bishop* is beautified as *Christopher Bishop*.

## 4. RESULTS

### 4.1 Experiments

1598 articles were collected from 110 different RSS feeds. The crawling system described in section 3.2 was used to collect them. RSS feeds were chosen according potential scenarios such as: items without images, items with post excerpts, items without categories, etc. Besides that, the sources of the RSS feeds are: online newspapers, blogs, magazines, entertainment sites and others. The variety of sources ensures that the proposed system works regardless the topic or source of the item.

The following attributes were extracted when possible for each item: title, link, content, author, published date and category. An analysis of the dataset was carried out before and after applying our system to it as seen in Table 1. It is necessary to point out that the attributes: title, link and date were always present in the items. Consequently, they always took value 1 when computing IQ explained in section 3.1.

Different methods were used to analyse the rest of the attributes for each item. HTML tags were removed from the content using a parser library called Beautiful Soup [Beautiful Soup Core Development Team (2004)].  Hence, a clean text was obtained and used to calculate the average content length of the items in the dataset. After that, the content (formatted as HTML) was parsed in order to extract images and calculate the total number of items with at least one image. Then, the total number of items with at least one category is calculated. Finally, the total number of items that have the author attribute is calculated. The method to extract categories and author is described in section 3.2.1.

### 4.2 Primary Results

The proposed system for improving the data quality of RSS feeds was applied to the initial dataset. Then, the resulting data set was analyzed again using the methods mentioned in the previous section. Both analysis are contrasted in Table 1 and Fig. 4.

| Attributes | Before | After |
| --- | --- | --- |
| Average Content Length (#chars) | 388.155 | 3211.082 |
| Articles with images/thumbnails | 130 | 1595 |
| Article with categories | 1219 | 1541 |
| Articles with Author | 1411 | 1437 |
| Average Item Quality | 39.98% | 95.62% |

Table 1. Content Analysis.

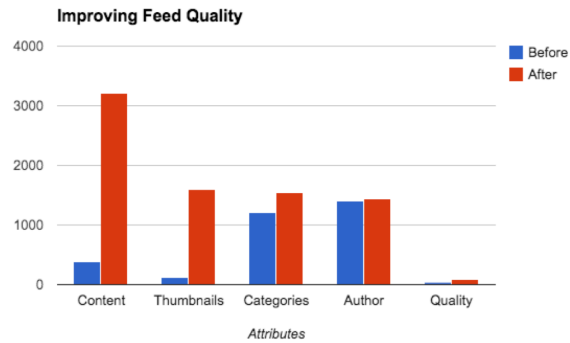

Fig. 4. Improving Feed Quality Graph.

In overall, the system has improved the average item quality from 39,98% to 95,62%. Therefore, it is a strong candidate for improving the quality of RSS feeds and transforming unstructured news pages into structured data.

For images, 1465 items did not have an image. 1356 out of 1465 images were extracted from the original web page where the item was located. The rest of images (109) were extracted from a public image database as described in section 3.5. In addition, a few images from the public image database did not match with the keywords provided. This opens a new research path to assess the quality of images matched with keywords from public image databases.

For categories, 376 items did not have categories assigned. 322 out of 376 categories were predicted with a confidence above 80%. 54 items were not categorised because the confidence was below 80%.

For authors, 184 items did not have the attribute author. 26 out of 184 authors were found using the proposed system.

Finally, a visual test was carried out using a popular feed aggregation app called Feedly [Feedly 2004]. A RSS feed was visualised before and after applying the proposed system as shown in Figure 5.

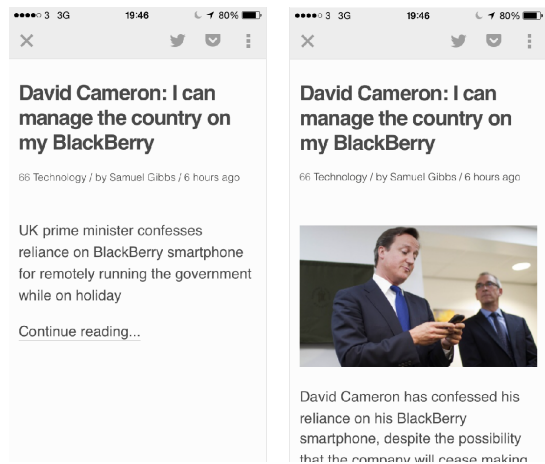

Fig. 5. Before and after RSS feed visualization.

## 5. CONCLUSIONS AND RECOMMENDATIONS

This paper described a new automated system for improving the quality of RSS feeds and transforming unstructured news pages into structured data.

The system efficiently collects the HTML pages where each feed item is located using its link attribute. Then it identifies where the main content is located, which is the region that has the highest number of characters per tag and are consecutive. It extracts the main text corpus and the relevant keywords. Following that, the most relevant image. Finally, the item is categorized using an OVR learning algorithm.

The system showed an item quality improvement from 39.98% to 95.62%. Therefore, it can be used by different applications that use RSS feeds such as: news aggregation sites (Flipboard, Stumbleupon or Google News), RSS readers (Feedly or Digg Reader) or content distribution platforms (Outbrain or Taboola).

The study can be extended by using natural language processing methods to extract relevant attributes such as: entities, quotes and locations. This would provide a better context to the news. The classification algorithm can be improved by adding more categories and subcategories, which requires collecting more labeled content from different publishers. Furthermore, it is strongly recommended to research a system for extracting media such as: videos, podcasts and infographics from pages.